\documentclass[aps,prl,reprint, 10pt,superscriptaddress,showpacs,amssymb,amsfonts,onecolumn,notitlepage,twocolumn]{revtex4-2}
\usepackage{amsmath,amssymb,amsthm,amscd,latexsym,epsfig,bm}

\usepackage[colorlinks,bookmarks=false,citecolor=blue,linkcolor=red,urlcolor=blue]{hyperref}
\usepackage{verbatim}

\usepackage{booktabs}
\usepackage{longtable}
\usepackage{amstext}
\usepackage{array}
\usepackage{multirow}
\usepackage[dvipsnames]{xcolor}
\usepackage{tabularx}
\usepackage{enumerate}   

\newcolumntype{L}{>{$}l<{$}} 
\newcolumntype{C}{>{$}c<{$}} 

\def\ket#1{\vert #1 \rangle}
\def\bra#1{\langle #1 \vert}

\def\f2{{\mathbb F}_2}

\theoremstyle{plain}
\theoremstyle{plain}

\providecommand{\theoremname}{Theorem}
\providecommand{\theoremtextname}{Theorem}

\theoremstyle{plain}
\providecommand{\propositionname}{Proposition}

\newcommand{\nc}{\newcommand}
\nc{\be}{\begin{equation}} \nc{\ee}{\end{equation}}
\nc{\bea}{\begin{eqnarray}} \nc{\eea}{\end{eqnarray}}
\nc{\bean}{\begin{eqnarray*}} \nc{\eean}{\end{eqnarray*}}
\nc{\dg}{\dagger}
\nc{\ua}{\uparrow} \nc{\da}{\downarrow}

\usepackage{color}
\usepackage[dvipsnames]{xcolor}
\usepackage{dsfont}

\newcommand{\bsl}[1]{\boldsymbol{#1}}

\renewcommand{\mod}{\,\mathrm{mod}\,}

\newcommand{\ii}{\mathrm{i}}

\newcommand{\dsZ}{\mathbb{Z}}

\newcommand{\eqnref}[1]{Eq.\,\eqref{#1}}

\newcommand{\eq}[1]{\begin{equation} #1 \end{equation}}

\newcommand{\eqa}[1]{\begin{align}\begin{split} #1 \end{split}\end{align}}

\let\oldAA\AA
\renewcommand{\AA}{\text{\normalfont\oldAA}}

\newcommand{\cc}{\mathcal{K}}

\renewcommand{\P}{\mathcal{P}}
\newcommand{\PT}{\mathcal{PT}}

\newcommand{\N}{\mathcal{N}}

\newcommand{\lcm}{\mathrm{lcm}}

\usepackage{ulem}

\begin{document}

\title{Classification of Interacting Dirac Semimetals}
\author{Sheng-Jie Huang}
\affiliation{Max Planck Institute for the Physics of Complex Systems, N{\"o}thnitzer Str. 38, 01187 Dresden, Germany}
\affiliation{Joint Quantum Institute, Department of Physics, University of Maryland, College Park, Maryland 20742-4111, USA}
\affiliation{Condensed Matter Theory Center, Department of Physics, University of Maryland, College Park, Maryland 20742-4111, USA}
\author{Jiabin Yu}
\affiliation{Condensed Matter Theory Center, Department of Physics, University of Maryland, College Park, Maryland 20742-4111, USA}
\affiliation{Department of Physics, Princeton University, Princeton, NJ 08544, USA}
\author{Rui-Xing Zhang}
\email{ruixing@utk.edu}
\affiliation{Department of Physics and Astronomy, University of Tennessee, Knoxville, Tennessee 37996, USA}
\affiliation{Department of Materials Science and Engineering, University of Tennessee, Knoxville, Tennessee 37996, USA}
\affiliation{Institute for Advanced Materials and Manufacturing, University of Tennessee, Knoxville, Tennessee 37920, USA}
\date{\today}

\begin{abstract} 
    Topological band theory predicts a $\mathbb{Z}$ classification of three-dimensional (3D) Dirac semimetals (DSMs) at the single-particle level. Namely, an arbitrary number of identical bulk Dirac nodes will always remain locally stable and gapless in the single-particle band spectrum, as long as the protecting symmetry is preserved. In this work, we find that this single-particle classification for $C_n$-symmetric DSMs will break down to $\mathbb{Z}_{n/\text{gcd}(2,n)}$ in the presence of symmetry-preserving electron interactions. Our theory is based on a dimensional reduction strategy which reduces a 3D Dirac fermions to 1D building blocks, i.e., vortex-line modes, while respecting all the key symmetries. Using bosonization technique, we find that there exists a minimal number $N=n/\text{gcd}(2,n)$ such that the collection of vortex-line modes in $N$ copies of DSMs can be symmetrically eliminated via four-fermion interactions. While this gapping mechanism does not have any free-fermion counterpart, it yields an intuitive ``electron-trion coupling" picture. By developing a topological field theory for DSMs and further checking the anomaly-free condition, we independently arrive at the same classification results. Our theory paves the way for understanding topological crystalline semimetallic phases in the strongly correlated regime.  
\end{abstract}

\maketitle

{\it Introduction} - Semimetallic crystalline solids with vanishing density of states near the Fermi level usually carry nontrivial topological properties~\cite{Burkov2016TSM,Yan2017WSM,Bernevig2018TSM,Vishwanath2018RMPTSM}. Dirac semimetal (DSM) is such an example with 4-fold degenerate point nodes in the bulk energy band spectrum, whose nodal quasiparticles resemble massless Dirac fermions in three dimensions (3D)~\cite{Young2012,Wang2012,Wang2013,Liu2014,Neupane2014,Yang2014,yang2015DSM,zhang2020DiracSC}. The Dirac points have symmetry-based $\mathbb{Z}$ topological indices~\cite{yang2015DSM,zhang2020DiracSC}, and the manifestation of the topological indices lies in their exotic boundary modes (i.e., Fermi arc states). In particular, unlike doubly degenerate Weyl points, stabilizing Dirac node (and the definition of the topological indices) in a 3D crystal will necessarily require the crystalline symmetries. Thus far, experimental realizations of DSM phase have been achieved in a plethora of quantum materials, including Na$_3$Bi, Cd$_3$As$_2$, etc.~\cite{Liu2014,Neupane2014}, which have been attracting great research attentions. 

The ubiquitous electron correlations in quantum materials, however, could prevent band theory from faithfully descriping real-world crystalline semimetals. For gapped topological systems, it has been well established that interaction effects can enable new topological phase that is free-fermion impossible, e.g., the anomalous surface topological orders in interacting topological insulators. Besdies, electron correlations can also trigger new topological equivalence relations between phases that are distinct in the non-interacting limit, qualitatively modifying the topological classification. Fidkowski and Kitaev first pointed out the $\mathbb{Z}\rightarrow \mathbb{Z}_8$ classification reduction for 1D class BDI superconductors~\cite{Fidkowski2010}. Similar reduction relations were later established for various crystalline topological insulators and superconductors~\cite{Yao2013,Isobe2015,Song2017,Song2017reduction,Aksoy2021}. Nonetheless, unlike their gapped cousins, little progress has been made towards understanding correlated gapless topological matters, where, in particular, their classification schemes have remained a fundamentally important open question.   

\begin{table}[t]
	\centering
	\includegraphics[width=1\linewidth]{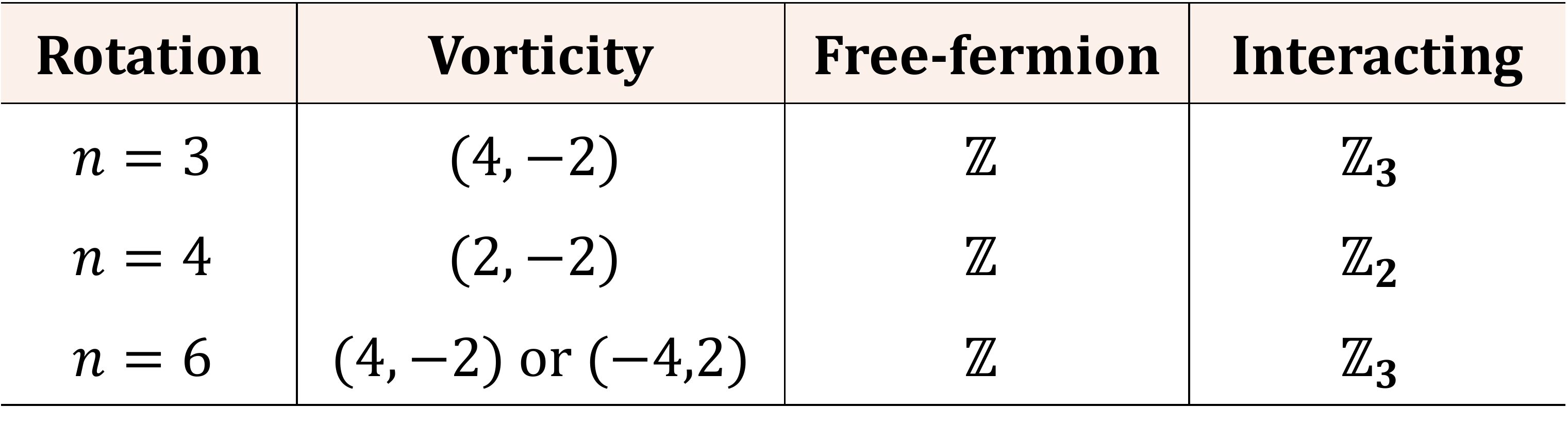}
	\caption{
	Classifications of 3D Dirac semimetals with $C_n$ rotation symmetries. ${\cal V}_n = (\nu_\text{v}, \nu_\text{av})$ denotes the minimal vorticities for vortex and antivortex, and all DSM$_n$s admit a $\mathbb{Z}$ classification for a fixed choice of symmetry representation (labelled by $l$  in the main text) in the free-fermion limit. The interacting classification for DSM$_n$ is given by $\mathbb{Z}_{n/\text{gcd}(2,n)}$, as derived from the bosonization theory and the quantum anomaly analysis. 		
	}	
	\label{table1}
\end{table}

\begin{figure*}[t]
	\centering
	\includegraphics[width=1\linewidth]{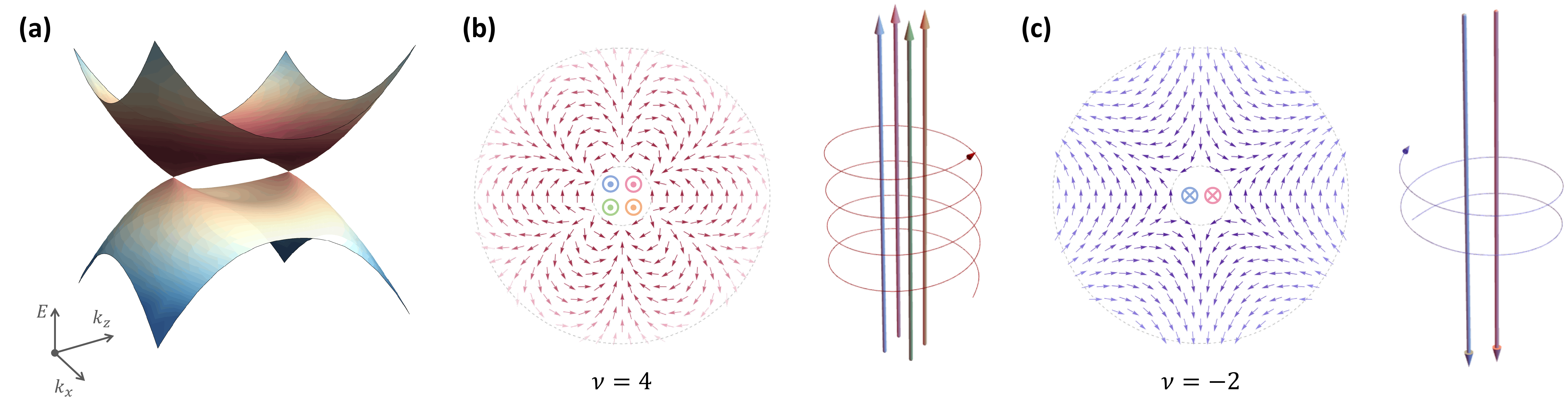}
	\caption{
		Dimensional reduction of DSM. For a DSM$_3$ with a pair of Dirac points is shown in (a), the minimal vorticities for a vortex (anti-vortex) mass is $\nu=4$ ($\nu=-2$), as shown in (b) and (c). The vortex will trap $|\nu|$ 1D gapless vortex-line modes at each valley for DSM$_3$, whose propagating direction depends on sgn$(\nu)$. Only vortex modes at the $v=+$ valley are shown here.
	}	
	\label{fig1}
\end{figure*}

In this work, we uncover a $\mathbb{Z}_{n/\text{gcd}(2,n)}$ classification for $C_n$-symmetric correlated spinful DSMs by extending the frameworks of ``topological crystals"~\cite{Isobe2015,Song2017,Huang2017,Shiozaki2018,Song2019,Else2019defect,Song2020real} and crystalline gauge fields~\cite{Else2018, Nissinen2018tetrads, nissinen2020field, Song2021polarization, Gioia2021, Naren2021,Julian2021,Huang2022,Julian2022} to semimetallic phases. Compared with the $\mathbb{Z}$-classification of free-fermion DSMs, this new classification reduction is purely interaction driven. A key step of our framework is the successful dimensional reduction for DSMs, with which 3D Dirac fermions can be symmetrically ``reduced" into 1D gapless fermionic {\it vortes-line} modes. This reduction procedure has been the main approach for classifying crystalline symmetry protected topological phases~\cite{Isobe2015,Song2017,Huang2017,Song2019}. As we apply it to DSMs, it well-positions us to identifying the lower-dimensional building blocks and to exploring correlation effects of the residue 1D vortex-line modes with the Luttinger liquid theory. In particular, we find that four-fermion interactions can gap out a collection of vortex-line modes without introducing any form of symmetry breaking, if the vortex-line modes originate from $N=n/\text{gcd}(2,n)$ copies of identical DSMs. This concludes our classification for correlated DSMs and the explicit gapping process can be understood through a {\it electron-trion coupling} picture. We further develop a topological field theory for DSMs and derive the $\mathbb{Z}_{n/\text{gcd}(2,n)}$ classification by checking the anomaly-free condition. Relation between our theory and Lieb-Schultz-Mattis constraint are also discussed.

{\it Band Theory of Dirac Semimetals} - In a 3D $C_n$-symmetric spinful crystal with both inversion symmetry ${\cal P}$ and time-reversal symmetry (TRS) $\Theta$, all energy bands must pair up to form degenerate band doublets. Along the $C_n$-invariant $k_z$ axis, two bands constituting a doublet are labeled by their $\hat{z}$-component angular momenta $\{J,-J\}$ with $J$ being an half-integer. When two doublets of bands cross at a generic $k_z=k_0$, they form a stable 4-fold degenerate Dirac node, when they carry  inequivalent angular momenta with $|J|\neq |J'|$ mod $n$. Assuming $k_0$ is not a time-reversal invariant momentum, TRS requires the presence of a partner Dirac node at $k_z=-k_0$ formed by the same set of bands, a manifestation of fermion doubling theorem. 
The topological stability of a Dirac point is guaranteed by a $\mathbb{Z}$-type symmetry charge ${\cal Q}_{J}$~\cite{yang2015DSM,zhang2020DiracSC}. Respecting the symmetries, a Dirac point can develop a gap only when merging wth another oppositely-charged Dirac point at the same momentum. Note that the key symmetries involving in the protection of a stable Dirac point are $U(1)$ charge conservation, $C_{n}$-rotation, inversion-time-reversal ${\cal P}\Theta$ and a lattice translation along the rotation axis.

Without loss of generality, we assume all Dirac points in a $C_n$-symmetric DSM (denoted as DSM$_n$) sitting at $k_z=\pm k_0$, i.e., the two ``valleys" labeled by an index $v=\pm$. The $\hat{z}$-directional lattice translation symmetry ${\cal T}_z$ promotes the valley index $v$ to be a good quantum number to mark the low-energy Dirac fermions. Meanwhile, ${\cal T}_z$ acts on the fermions as a valley $U(1)$ rotation $\exp{(i\alpha_{z} \varphi)}$, where the phase angle $\varphi$ can be promoted to take continuous values at low energies. At each valley $v$, the minimal Hamiltonian for a single Dirac point reads
\begin{equation}
	H_0^{(v)} ({\bf k})=k_x\gamma_1 + k_y \gamma_2 + v k_z \gamma_5 + {\cal O}(k^2),
	\label{eq:3dDiracH}
\end{equation}
where we have defined the $\gamma$ matrices
$ \gamma_1=\tau_3 \otimes \sigma_1,
\gamma_2=\tau_3 \otimes \sigma_2,
\gamma_3=\tau_1 \otimes \sigma_0,
\gamma_4=\tau_2 \otimes \sigma_0,
\gamma_5=\tau_3 \otimes \sigma_3 $
and 
$\gamma_{jk}=\frac{1}{2i}[\gamma_j,\gamma_k],\ \forall\ j<k$. Here, $k_z$ is an effective crystal momentum relative to the valley. The inversion symmetry and TRS are represented by $\mathcal{P} \doteq \alpha_{x} \otimes \gamma_{5}$ and $\Theta \doteq i\alpha_{x} \otimes \gamma_{13} \cc$, respectively, where $\alpha_{x,y,z}$ are Pauli matrices for the valley degree of freedom (d.o.f.) and $\cc$ the complex conjugate~\footnote{Most existing DSM materials, such as $\text{Na}_{3}\text{Bi}$ and $\text{Cd}_{3}\text{As}_{2}$, belong to this representation of inversion operator. The other possible choice is $\mathcal{P} \doteq s_{x} \otimes \gamma_{0}$. These choices won't affect the classification.}. Under this basis choice, $C_n$ has a diagonal matrix representation $C_n\doteq\exp(i \frac{2\pi}{n}J)$, where $J=\text{diag}(l+\frac{1}{2}, l-\frac{1}{2}, -l+\frac{1}{2}, -l-\frac{1}{2})$ and $l<n/2$ is a positive integer. One can again check that a single-particle Dirac mass for \eqnref{eq:3dDiracH} must break either $C_n$ or ${\cal T}_z$~\cite{zhang2016nematic}. 

{\it Dimensional Reduction} - Although the gaplessness of free-fermion DSMs is symmetrically irremovable, however, we find it's possible to reduce the dimensionality of gapless d.o.f. from 3D to 1D, while respecting all the key symmetries. This {\it dimensional reduction} procedure for Dirac fermions is motivated by the observation that $(\gamma_3, \gamma_4)$ form a vector representation of $C_n$. We thus define $\gamma_{\pm} = \frac{1}{2}(\gamma_3\pm i\gamma_4)$ and find that
\begin{equation}
    C_n \gamma_{\pm} C_n^{-1} = e^{\pm i \frac{4\pi}{n} l} \gamma_{\pm}.
\end{equation}
Namely, $\gamma_{\pm}$ carry angular momenta $J=\pm 2l$, respectively. As $\gamma_{3/4}$ breaks $C_n$ explicitly, we need to couple $\gamma_{\pm}$ to some spatial-dependent functions with proper angular momenta to preserve $C_n$. This motivate us to consider a vortex mass term for the Dirac point at valley $v$, 
\begin{equation}
    H_\nu^{(v)} (r,\theta) = v m_0(r)[ e^{i\nu\theta}\gamma_+ +  e^{-i\nu\theta}\gamma_-],
\label{eq:vmass}
\end{equation}
where $(r,\theta)$ are the in-plane polar coordinates with $r=\sqrt{x^2+y^2}$ and $\theta=\tan^{-1}(y/x)$. We choose $m_0(r)=m_0$ to be spatially uniform. To respect both $C_n$ and $\mathcal{P}\Theta$, the vorticity $\nu$ of the vortex mass must satisfy
\begin{eqnarray}
 \nu = pn-2 l
\label{eq:ne_cond}
\end{eqnarray}
with $ \nu\in 2\mathbb{Z}$ and $p \in \mathbb{Z}$. Importantly, $H_{\nu}$ is unable to eliminate all gapless d.o.f., as we analytically derived the gapless vortex-line modes dispersing along $k_z$ in the presence of $H_{\nu}$. (See Supplemental Materials (SM)~\cite{SM} for details.) The vortex-line modes are found to carry a half-integer-valued $z$-component angular momentum $J$ (mod $n$), with 
\begin{eqnarray}
	&& l-\frac{1}{2}<J<\nu+l+\frac{1}{2},\ \ \text{for }\nu>0, \nonumber \\
	&& \nu+l-\frac{1}{2} < J < l+\frac{1}{2},\ \ \text{for } \nu<0.
\end{eqnarray}
One can check that the number of zero modes is given by $|\nu| = |pn-2 l|$. Using perturbation theory, we find that the vortex-line mode at valley $v$ is locally chiral with $E(k_{z}) = \text{sgn}(\nu v)k_{z}$ and an angular momentum label $J$. Due to this {\it chirality-valley locking} effect, in the SM~\cite{SM}, we find it impossible to symmetrically gap out the $2|\nu|$ chiral modes in a ``minimal" DSM$_n$ with a pair of Dirac nodes, even when interaction effects are considered. This completes our dimensional reduction procedure, where we have successfully ``reduced" 3D Dirac physics into 1D vortex-line modes in a fully symmetric manner.  The vortex-line modes are thus the {\it 1D building blocks} for general 3D DSMs.

{\it Classification \& Interaction Effects} - Classifying DSM$_n$s is equivalent to studying the stability problem of Dirac points. Consider $N\in\mathbb{Z}_{>0}$ identical copies of minimal DSM$_n$s, each featuring a pair of Dirac nodes at $\pm k_0$. This DSM$_n$ phase will admit a $\mathbb{Z}_N$ classification, if there exists a minimal $N$ such that all Dirac nodes can be gapped out thoroughly without introducing (i) explicit or spontaneous symmetry breaking; or (ii) topological order. Otherwise, the DSM$_n$ is $\mathbb{Z}$ classified. 

We now attack this classification problem with the help of dimensional reduction procedure. Take DSM$_3$ as an example, where the basis index $l=1$ is the only choice respecting $0<l<n/2$. We define ${\cal V}_n = (\nu_\text{v}, \nu_\text{av})$ to denote the minimal symmetry-allowed vorticity for vortex ($\nu_\text{v}>0$) and anti-vortex ($\nu_\text{av}<0$) for DSM$_n$. \eqnref{eq:ne_cond} immediately implies that ${\cal V}_3 = (4,-2)$ and a summary of ${\cal V}_n$ can be found in Table.~\ref{table1}. We further denote $\psi_{v,J,R/L}$ as a fermion operator that annihilates a right/left chiral vortex-line mode with an angular momenta $J$ at valley $v$. In particular, a $\nu=4$ vortex harbors four right-moving vortex-line modes $\Psi_{+, \nu=4}^{(3)} = (\psi_{+,\frac{3}{2},R}, \psi_{+,\frac{1}{2},R}, \psi_{+,-\frac{1}{2},R}, \psi_{+,-\frac{3}{2},R})$ and a $\nu=-2$ anti-vortex traps two right movers $\Psi_{+, \nu=-2}^{(3)} = (\psi_{+,\frac{1}{2},L}, \psi_{+,-\frac{1}{2},L})$ at the $+k_0$ valley. The $-k_0$ valley states are TRS-related with $\Psi_{-,\nu}^{(3)} = \Psi_{+, \nu}^{(3)}(R \leftrightarrow L)$. At the free-fermion level, it is straightforward to check the inability to find a combination of vortices and antivortices that can gap out the vortex-line modes, a manifestation of the  $\mathbb{Z}$ classification of non-interacting DSMs. 

When interaction effects are considered, the 1D vortex-line modes are readily described by the Luttinger liquid theory. Using the bosonization technique, we show that for DSM$_n$, a group of vortex-line modes can always be symmetrically gapped out if the net vorticity vanishes. (See SM~\cite{SM} for details.) As a result, for $N=3$ copies of DSM$_3$, one can decorate one DSM$_3$ copy with a $\nu=4$ vortex and the other two with a $\nu=-2$ vortex each. Then the cancellation of vorticities directly implies the symmetry-preserving gapping of the vortex-line modes, immediately leading to a $\mathbb{Z}_3$ classification of DSM$_3$. Namely, DSM$_3$s with $N$ pairs of Dirac points can be gapped out symmetrically by interactions, as long as $N$ is an integer multiple of $3$.
 
For DSM$_4$, we have ${\cal V}_4 = (2, -2)$. Therefore, two identical copies of vortex-decorated DSM$_4$, one with a vortex and another with an anti-vortex, will enable a symmetric gapping, further leading to the $\mathbb{Z}_2$ classification for DSM$_4$. For DSM$_6$, there exist two different basis choices with $l=1,2$, which leads to ${\cal V}_6^{(l=1)} = (4,-2)$ and ${\cal V}_6^{(l=2)} = (2,-4)$, respectively. For both scenarios, we will need three copies of DSM$_6$ to realize the cancellation of vorticity, similar to the case in DSM$_3$. We thus conclude that DSM$_6$ generally admits a $\mathbb{Z}_3$ classification in the correlated limit. A summary of classification of DSM$_n$ in both free-fermion and interacting limits can be found in Table.~\ref{table1}.
 
{\it Electron-Trion Coupling} - We now provide an alternative yet intuitive ``trion" picture to understand how a symmetric many-body gap will naturally arise in DSM$_n$. Let us focus on DSM$_3$, where the low-energy vortex-line modes involved in a three-vortex configuration are one $\Psi_{+, \nu=4}^{(3)}$ and two $\Psi_{+, \nu=-2}^{(3)}$ at valley $+k_0$, as well as their TRS partners. We consider a four-fermion term 
\begin{eqnarray}
	H_\text{ET}=\psi^\dagger_{+,\frac{3}{2},R} \Phi_{+,\frac{3}{2},L},
\end{eqnarray}
where $\Phi_{+,\frac{3}{2},L} = \psi^\dagger_{-,-\frac{1}{2},R}\psi_{+,-\frac{1}{2},L}\psi_{-,\frac{3}{2},L}$ annihilates a left-moving three-particle excitation (i.e., a trion~\cite{Berg2022GappingFragile}). Clearly, the trion carries exactly the same set of symmetry indices as that of $\psi^\dagger_{+,\frac{3}{2},R}$, so that the above electron-trion interaction is symmetry-preserving and will unambiguously gap out the vortex mode generated by $\psi^\dagger_{+,\frac{3}{2},R}$.  

When the net vorticity of vortex masses vanishes, the collection of vortex-line modes are non-chiral as a whole. This ensures that we can write down a complete set of electron-trion terms to gap out all fermionic d.o.f., which is further supported by an explicit bosonization analysis in the SM~\cite{SM}. Since $\psi_{+,\frac{3}{2},L}$ is absent in the vortex-line mode basis, the trion $\Phi_{+,\frac{3}{2},L}$ does NOT have any single-fermion counterpart that share both the same symmetry labels and the chirality. Therefore, the above gapping process is only possible via many-body interactions.     

{\it Topological Field Theory \& Quantum Anomaly} - Here we derive the classification from the effective field theory of the Dirac semimetal. A brief review of the crystalline gauge fields and the relation to spatially dependent mass terms is given in SM~\cite{SM} (See also Ref.~\cite{Else2018, Nissinen2018tetrads, nissinen2020field, Song2021polarization, Gioia2021, Naren2021,Julian2021,Huang2022,Julian2022}). We couple the theory to the crystalline gauge fields through adding the spatially dependent mass terms such as the vortex mass discussed above, which is the approach introduced in Ref.~\cite{Huang2022}.

To explicitly couple the 3D Dirac semimetal to a translation gauge field in the z-direction, we add the following spatially dependent mass term (in addition to the vortex mass \eqnref{eq:vmass}):
\begin{equation}
   H_z = m'e^{i \kappa \phi(z) \gamma_{5}},
\label{eq:zmass}
\end{equation}
where $\kappa = 2k_{0}/2\pi$, and $2k_{0}$ is the momentum separation of the Dirac points at the two valleys. Translation in the $z$-direction acts on $\phi$ by shifting $\phi \rightarrow \phi + 2\pi \mathbb{Z}$.

After coupling the Dirac semimetal to a $U(1)$ gauge field in the presence of these spatially dependent mass terms and integrating out the Dirac fermions,
we obtain the unquantized topological term:
\begin{equation}
    S = \kappa \int A \wedge E_{z} \wedge d\omega^{(n)},
\label{eq:Stopocn}
\end{equation}
where $E_{z} = d\phi/2\pi$ are the translation gauge field and $\omega^{(n)} = d n_{\theta}/2\pi$ with $n_{\theta} = \nu \theta$ is the rotational gauge field. Note that $E_{z}$ and $\omega^{(n)}$ are in the non-trivial cohomology class in $H^{1}(B \Gamma,\mathbb{Z})$ and $H^{1}(BC_{n},\mathbb{Z})$, respectively (see SM~\cite{SM} for more details). For a single DSM, one can show that the 2-form $d\omega^{(n)}$ must satisfy 
\begin{equation}
    \int_{M_{xy}} d\omega^{(n)} = \nu \ \text{mod} \ n,
\end{equation}
where $M_{xy}$ is a $xy$-plane. The topological term \eqnref{eq:Stopocn} is not gauge invariant since $\kappa$ is fractional, which signals the quantum anomaly. If we integrate over the $xy$-plane, we find the anomaly localized and there is a 1D filling anomaly \cite{Gioia2021} for the vortex-line modes.


Now we show how to derive the classification of the Dirac semimetals from the field theory perspective. When we have $N$ copies of the Dirac semimetal, the 2-form $d \omega^{(n)}$ satisfies 
\begin{equation}
     \int_{M_{xy}} d\omega^{(n)} = N(\nu\ \text{mod}\ n) \ \text{mod}\ n =  N\nu \ \text{mod} \ n. 
\label{eq:rotocond}
\end{equation}
When $\int d\omega^{(n)} = 0$ mod $n$, the rotation gauge field is in the trivial cohomology class and the anomaly in \eqnref{eq:Stopocn} vanishes, which implies that the Dirac semimetals can be gapped out while preserving the symmetry. Note that $\nu \in 2\mathbb{Z}$ due to $\mathcal{PT}$ symmetry. Therefore, the classification can be obtained by solving the equation $N\nu =0$ mod $n$. 
As elaborated in SM~\cite{SM}, the solution to the equation is $N \in  \frac{n}{\text{gcd}(|\nu|,n)} \dsZ$.
%
Hence, interacting classification of the Dirac semimetals is given by $\mathbb{Z}_{\frac{n}{\text{gcd}(|\nu|,n)}}$ with the minimum allowed value of $|\nu|=2$, which is consistent with the result obtained by the direct analysis of the dimension reduction summarized in Table.~\ref{table1}~\footnote{We note that these classification is valid when $\kappa$ (hence the momentum separation of the Dirac points) doesn't belong to the set of exceptional values: $\kappa_{ex} = p/(\nu+qn)$, where $p$ and $q$ are integers. When $\kappa = \kappa_{ex}$, we can shift the rational gague field $\omega^{(n)}$ by a coboundary such that $\int d \omega^{(n)} = \nu + q n$, and the anomaly is in fact trivial. This implies that the Dirac points can be gapped out symmetrically by strong interactions. Note that shifting the quantization condition of $d \omega^{(n)}$ might require introducing additional bands in practice. Similar conclusions of the exceptional values are obtained in Ref.~\onlinecite{Gioia2021} for different topological semimetals.}.

{\it Discussions -} We have shown that correlated 3D spinful Dirac semimetals yield a qualitatively different $\mathbb{Z}_{\frac{n}{\text{gcd}(|\nu|,n)}}$ classification from that of their free-fermion counterparts (i.e., $\mathbb{Z}$). Our classification theory is based on identifying the low-dimensional building blocks for DSMs by coupling the Dirac fermions to fully symmetric vortex masses. Remarkably, the 1D blocks are gapless vortex-line modes that are symmetrically irremovable at the free-fermion level. This dimensional reduction technique enables both an exact bosonization approach to deal with electron correlation effects and a topological field theory construction to analyze quantum anomaly of DSMs, both of which lead to the same classification result for interacting DSMs. We note that while our theory is physically reasonable, a mathematical rigorous proof of the classification is still an open question.

When there are $N<n/\text{gcd}(2,n)$ copies of DSM$_n$, the system is a symmetry-protected correlated gapless phase. To generate mass for the Dirac fermions, one will then have to either break the symmetries~\cite{zhang2016nematic} or introduce electron fractionalization~\cite{Teo2019,Wang2020}. In other words, for $N<n/\text{gcd}(2,n)$, our classification suggests the absence of any symmetric featureless insulators, which manifests as a generalized Lieb-Schultz-Mattis constraint. When $N= n/\text{gcd}(2,n)$, the system is free from the above constraint and can always be turned into a featurelessly insulator, as shown by our explicit construction. This implies the possibility of having a direct symmetry mass generation (SMG) transition towards a featurelessly insulator~\cite{Wang2022smg}.

Our theory will shed new light on comprehending the topological nature of correlated semimetals, and in particular their quantum anomaly phenomena. Our results are timely, especially given the rapidly growing list of correlated semimetals candidates, including Ce$_3$Bi$_4$Pd$_3$~\cite{dzsaber2017kondo}, Ce$_2$Au$_3$In$_5$~\cite{chen2022topological}, CaIrO$_3$~\cite{fujioka2019strong}, and SrNbO$_3$~\cite{ok2021correlated}, etc. Finally, we highlight that the methodology used in this work is general and directly applicable to other correlated crystalline semimetals. We are thus confident that our framework will pave the way for a unified, beyond-band theory for symmetry-protected semimetallic physics.

{\it Acknowledgements} - SJH and RXZ are grateful to the 2018 Boulder Summer School, where this work was initiated. SJH acknowledges the support from a JQI Postdoctoral Fellowship. JY acknowledges the support by the Gordon and Betty Moore Foundation through the EPiQS Initiative. RXZ acknowledges the start-up fund of the University of Tennessee. The work done at the University of Maryland is supported by the Laboratory of Physical Sciences.

\bibliography{semimetal}

\newpage

\appendix

\section{Zero mode solutions of the vortex-line modes}
\label{app:zero}
To solve the zero mode bound stats at the vortex center, we propose a trial wavefunction
\begin{align}
\Psi=[f_1(r)e^{i(J-l-\frac{1}{2})\theta}, f_2(r)e^{i(J-nl_2-l+\frac{1}{2})\theta}, 
\\
f_3(r)e^{i(J-nl_3+l-\frac{1}{2})\theta},f_4(r)e^{i(J-nl_4+l+\frac{1}{2})\theta}]^T.
\end{align}
where $J$ is a half-odd integer and $l_2,l_3,l_4\in\mathbb{Z}$. Plugging into the zero energy equation, we immediately arrive at the following equations:
\begin{align}
   & [i\partial_r+i\frac{J-nl_2-l+\frac{1}{2}}{r}]f_2(r)=mf_3(r)e^{in(l_2-l_3+p)\theta}, \\
   & [i\partial_r-i\frac{J-l-\frac{1}{2}}{r}]f_1(r)=mf_4(r)e^{in(p-l_4)\theta}, \\ 
   & mf_1(r)e^{in(l_4-p)\theta}=[-i\partial_r-i\frac{J-nl_4+l+\frac{1}{2}}{r}]f_4(r), \\ 
   & mf_2(r)e^{in(l_3-l_2-p)\theta}=[-i\partial_r+i\frac{J-nl_3+l-\frac{1}{2}}{r}]f_3(r).
\end{align}
It is instructive to integrate out the angular variable $\theta$ and we arrive at
\begin{align}
    & [i\partial_r+i\frac{J-nl_2-l+\frac{1}{2}}{r}]f_2(r)=mf_3(r)\delta_{l_2+p,l_3}, \\ 
& [i\partial_r-i\frac{J-l-\frac{1}{2}}{r}]f_1(r)=mf_4(r)\delta_{l_4,p}, \\ 
& mf_1(r)\delta_{l_4,p}=[-i\partial_r-i\frac{J-nl_4+l+\frac{1}{2}}{r}]f_4(r), \\ 
& mf_2(r)\delta_{l_3,l_2+p}=[-i\partial_r+i\frac{J-nl_3+l-\frac{1}{2}}{r}]f_3(r).
\end{align}

When the delta function in the above equations are evaluated to zero, it is easy to show that we can only have solutions of the form $f_i\sim x^{\alpha_i}$ which are not square integrable. Therefore, the physical solutions are possible only when the following conditions are satisfied,
\bea
l_3&=&l_2+p,\nonumber \\
l_4&=&p.
\eea

In this case, we can rewrite our trial wavefunction as
\bea
\begin{split}
\Psi=[f_1(r)e^{i(J-l-\frac{1}{2})\theta},f_2(r)e^{i(J'-l+\frac{1}{2})\theta},
\\
f_3(r)e^{i(J'-pn+l-\frac{1}{2})\theta},f_4(r)e^{i(J-pn+l+\frac{1}{2})\theta}]^T,
\end{split}
\eea
where $J'=J-nl_2$. Then the above equations become
\bea
&&[i\partial_r+i\frac{J'-l+\frac{1}{2}}{r}]f_2(r)=mf_3(r) \\ \label{Eq: E1}
&&[i\partial_r-i\frac{J-l-\frac{1}{2}}{r}]f_1(r)=mf_4(r) \\ \label{Eq: E2}
&&mf_1(r)=[-i\partial_r-i\frac{J-pn+l+\frac{1}{2}}{r}]f_4(r) \\ \label{Eq: E3}
&&mf_2(r)=[-i\partial_r+i\frac{J'-pn+l-\frac{1}{2}}{r}]f_3(r) \label{Eq: E4}
\eea
Plugging \eqnref{Eq: E2} into \eqnref{Eq: E3} and \eqnref{Eq: E4} into \eqnref{Eq: E1}, we arrive at 
\begin{widetext}
\bea
\left[ \partial_{\rho}^2+\frac{2 l+1-pn}{\rho}\partial_{\rho}-\frac{\rho^2+(J+l-pn-\frac{1}{2})(J-l-\frac{1}{2})}{\rho^2} \right] f_1(\rho)&=&0 \label{Eq: f1} 
\\  
\left[ \partial_{\rho}^2+\frac{n-2 l+p1}{\rho}\partial_{\rho}-\frac{\rho^2+(J'+l-pn-\frac{1}{2})(J'-l-\frac{1}{2})}{\rho^2} \right] f_3(\rho)&=&0 \label{Eq: f3}
\eea
\end{widetext}
where we have defined $\rho=|m|r$. \eqnref{Eq: f1} and \eqnref{Eq: f3} can be unified in a neat form:
\begin{align}
    [\partial_{\rho}^2+\frac{2\alpha+1}{\rho}\partial_{\rho}-\frac{\rho^2+(L-\frac{pn+1}{2})^2-\alpha^2}{\rho^2}]f(\rho)=0,
\label{Eq: Unified Equation}
\end{align}
where
\begin{itemize}
	\item \eqnref{Eq: f1}: $\alpha=l-\frac{n}{2}, L=J, f=f_1$;
	\item \eqnref{Eq: f3}: $\alpha=-(l-\frac{n}{2}), L=J', f=f_3$.
\end{itemize}
It is instructive to define 
\bea
\tilde{f}_{\alpha}=f\rho^{\alpha}.
\eea
which transforms \eqnref{Eq: Unified Equation} into the modified Bessel equation:
\bea
(\rho^2\partial_{\rho}^2+\rho\partial_{\rho}-[\rho^2+(L-\frac{pn+1}{2})^2])\tilde{f}_{\alpha}(\rho)=0.
\eea
The decay solution of the above equation is known as the modified Bessel function of the second kind,
\bea
\tilde{f}_{\alpha}(\rho)=cK_{L-\frac{pn+1}{2}}(\rho).
\eea
Therefore, we arrive at the following solutions for $f_1(\rho)$ and $f_3(\rho)$
\bea
f_1(\rho)&=&c_1\rho^{\frac{pn}{2}-l}K_{J-\frac{pn+1}{2}}(\rho) \nonumber \\
f_3(\rho)&=&c_3\rho^{-(\frac{pn}{2}-l)}K_{J'-\frac{pn+1}{2}}(\rho)
\eea
Meanwhile, from \eqnref{Eq: E2} and \eqnref{Eq: E4}, we arrive at
\bea
f_2(\rho) &=& ic_3\rho^{l-\frac{pn}{2}}K_{J'-\frac{pn-1}{2}}(\rho) \nonumber
\\
f_4(\rho) &=&-ic_1\rho^{-l+\frac{pn}{2}}K_{J-\frac{pn-1}{2}}(\rho).
\eea
where we have applied the recurrence relation of the modified Bessel functions:
\bea
(\partial_x+\frac{n}{x})K_n(x)&=&-K_{n-1}(x) \nonumber \\
(\partial_x-\frac{n}{x})K_n(x)&=&-K_{n+1}(x)
\eea
In summary, the radial part of the zero mode solutions are
\bea
f_1(\rho)&=&c_1\rho^{\frac{pn}{2}-l}K_{J-\frac{pn+1}{2}}(\rho) \nonumber \\
f_2(\rho)&=&ic_3\rho^{l-\frac{pn}{2}}K_{J'-\frac{pn-1}{2}}(\rho) \nonumber \\
f_3(\rho)&=&c_3\rho^{l-\frac{pn}{2}}K_{J'-\frac{pn+1}{2}}(\rho) \nonumber \\
f_4(\rho)&=&-ic_1\rho^{\frac{pn}{2}-l}K_{J-\frac{pn-1}{2}}(\rho)
\label{Eq: Zero mode general solutions}
\eea 

A physical wavefunction must be square integrable to be normalizable. The square integrability condition will put a strong constraint on \eqnref{Eq: Zero mode general solutions}. Since the modified Bessel function of the second kind is exponentially decaying for large $\rho$, we only need focus on the small $\rho$ behaviors for the solutions in \eqnref{Eq: Zero mode general solutions}. We note that 
\begin{itemize}
	\item If the solution $f(\rho)$ in \eqnref{Eq: Zero mode general solutions} satisfies $|f(\rho)|^2<\rho^{-2}$ as $\rho\rightarrow 0$, then $f(\rho)$ is square integrable.
\end{itemize}
By performing the Taylor expansion of $K_n(\rho)$ around $\rho=0$, we find the dominating contributions are given by
\bea
K_0(\rho)&=&-\log\rho+... \nonumber \\
K_n(\rho)&\sim&\rho^{-|n|}+...\ \ \ \forall n\neq0
\eea
Thus it is suggestive to discuss the following situations case by case. 

We first solve the case for $\nu = pn-2 l >0$. When $J=J'=\frac{pn+1}{2}$, we apply the square integrability condition to the zero mode solutions:
\begin{itemize}
	\item $|f_1(\rho)|^2 \sim \rho^{pn-2 l}|K_0(\rho)|^2 = \rho^{pn-2 l}(\log \rho)^2 \leq \rho^2(\log \rho)^2 \rightarrow 0$, as $\rho\rightarrow 0$. Thus, $f_1(\rho)$ is square integrable.
	\item $|f_2(\rho)|^2 \sim \rho^{2 l-pn}|K_1(\rho)|^2 = \rho^{2 l-pn-2} \geq \rho^{-4} > \rho^{-2}$, as $\rho\rightarrow 0$. Thus, $f_2(\rho)$ is NOT square integrable.
	\item $|f_3(\rho)|^2 \sim \rho^{2 l-pn}|K_0(\rho)|^2 = \rho^{2 l-pn}(\log \rho)^2 \geq \rho^{-2}(\log \rho)^2 > \rho^{-2}$, as $\rho\rightarrow 0$. Thus, $f_3(\rho)$ is NOT square integrable.
	\item $|f_4(\rho)|^2 \sim \rho^{pn-2 l}|K_1(\rho)|^2 = \rho^{pn-2 l-2} \rightarrow 0$, as $\rho\rightarrow 0$. Thus, $f_4(\rho)$ is square integrable.
\end{itemize}

When $J=J'=\frac{pn-1}{2}$, we apply the square integrability condition to the zero mode solutions:
\begin{itemize}
	\item $|f_1(\rho)|^2 \sim \rho^{pn-2 l}|K_{-1}(\rho)|^2 = \rho^{pn-2 l-2} \rightarrow 0$, as $\rho\rightarrow 0$. Thus, $f_1(\rho)$ is square integrable.
	\item $|f_2(\rho)|^2 \sim \rho^{2 l-pn}|K_0(\rho)|^2 = \rho^{2 l-pn}(\log \rho)^2 \geq \rho^{-2}(\log \rho)^2 > \rho^{-2}$, as $\rho\rightarrow 0$. Thus, $f_2(\rho)$ is NOT square integrable.
	\item $|f_3(\rho)|^2 \sim \rho^{2 l-pn}|K_{-1}(\rho)|^2 = \rho^{2 l-pn-2} \geq \rho^{-4} > \rho^{-2}$, as $\rho\rightarrow 0$. Thus, $f_3(\rho)$ is NOT square integrable.
	\item $|f_4(\rho)|^2 \sim \rho^{pn-2 l}|K_0(\rho)|^2 = \rho^{pn-2 l}(\log \rho)^2 \leq \rho^2(\log \rho)^2 \rightarrow 0$, as $\rho\rightarrow 0$. Thus, $f_4(\rho)$ is square integrable.
\end{itemize}

When $J=J'\neq \frac{pn+1}{2}$ and $J=J'\neq \frac{pn-1}{2}$, we find that
\begin{itemize}
	\item $|f_1(\rho)|^2\sim \rho^{pn-2 l-|2l-(pn+1)|}$: To make $|f_1(\rho)|^2<\rho^{-2}$ as $\rho\rightarrow 0$, we require
	\bea
	l-\frac{1}{2}<J<pn-l+\frac{3}{2}.
	\eea
	\item $|f_2(\rho)|^2 \sim \rho^{2 l-pn-|2l-(pn-1)|}$: To make $|f_2(\rho)|^2<\rho^{-2}$ as $\rho\rightarrow 0$, we require
	\bea
	0\leq |J-\frac{(pn-1)}{2}|<l-\frac{pn}{2}+1\leq 0,
	\eea
	which is impossible.
	\item $|f_3(\rho)|^2 \sim \rho^{2 l-pn-|2l-(pn+1)|}$: To make $|f_3(\rho)|^2<\rho^{-2}$ as $\rho\rightarrow 0$, we require
	\bea
	0\leq |J-\frac{(pn+1)}{2}|<l-\frac{pn}{2}+1\leq 0,
	\eea
	which is impossible.
	\item $|f_4(\rho)|^2\sim \rho^{pn-2 l-|2l-(pn-1)|}$: To make $|f_4(\rho)|^2<\rho^{-2}$ as $\rho\rightarrow 0$, we require
	\bea
	l-\frac{3}{2}<J<pn-l+\frac{1}{2}.
	\eea
\end{itemize}
Therefore, $c_2$ and $c_3$ have to be zero to make the zero mode wavefunction square integrable. In addition, we would require $l-\frac{1}{2}<J<pn-l+\frac{1}{2}$ to make $f_1(\rho)$ and $f_4(\rho)$ normalizable. To summarize, the zero mode wavefunction of the $Z_n$ vortex takes the following form:
\begin{align}
\Psi_{J,p,l}(\rho,\theta)=\frac{\rho^{\frac{pn}{2}-l}}{\sqrt{{\cal N}}}\begin{pmatrix}
	e^{(J-l-\frac{1}{2})\theta}K_{J-\frac{pn+1}{2}}(\rho) \\
	0\\
	0\\
	-ie^{(J-pn+l+\frac{1}{2})\theta}K_{J-\frac{pn-1}{2}}(\rho)
\end{pmatrix}
\end{align}
where ${\cal N}$ is the normalization factor. This solution is physical only when 
\bea
l-\frac{1}{2}<J<pn-l+\frac{1}{2},
\eea
provided that $\nu = pn-2 l >0$.

Now we solve the case for $\nu = pn-2 l <0$. When $J=J'=\frac{pn+1}{2}$, we apply the square integrability condition to the zero mode solutions:
\begin{itemize}
	\item $|f_1(\rho)|^2 \sim \rho^{-|\nu|}|K_0(\rho)|^2 = \rho^{-|\nu|}(\log \rho)^2 \geq \rho^{-2}(\log \rho)^2 > \rho^{-2}$, as $\rho\rightarrow 0$. Thus, $f_1(\rho)$ is NOT square integrable.
	
	\item $|f_2(\rho)|^2 \sim \rho^{|\nu|}|K_1(\rho)|^2 = \rho^{|\nu|-2} \leq \rho^{0} < \rho^{-2}$, as $\rho\rightarrow 0$. Thus, $f_2(\rho)$ is square integrable.
	
	\item $|f_3(\rho)|^2 \sim \rho^{|\nu|}|K_0(\rho)|^2 = \rho^{|\nu|}(\log \rho)^2 \leq \rho^{2}(\log \rho)^2  \rightarrow 0$, as $\rho\rightarrow 0$. Thus, $f_3(\rho)$ is square integrable.
	
	\item $|f_4(\rho)|^2 \sim \rho^{-|\nu|}|K_1(\rho)|^2 = \rho^{-|\nu|-2} \geq \rho^{-4} > \rho^{-2}$, as $\rho\rightarrow 0$. Thus, $f_4(\rho)$ is NOT square integrable.
\end{itemize}

When $J=J'=\frac{pn-1}{2}$, we apply the square integrability condition to the zero mode solutions:
\begin{itemize}
	\item $|f_1(\rho)|^2 \sim \rho^{-|\nu|}|K_{-1}(\rho)|^2 = \rho^{-|\nu|-2} \geq \rho^{-4} > \rho^{-2}$, as $\rho\rightarrow 0$. Thus, $f_1(\rho)$ is NOT square integrable.
	
	\item $|f_2(\rho)|^2 \sim \rho^{|\nu|}|K_0(\rho)|^2 = \rho^{|\nu|}(\log \rho)^{2} \leq \rho^{2}(\log \rho)^{2} \rightarrow 0$, as $\rho\rightarrow 0$. Thus, $f_2(\rho)$ is square integrable.
	
	\item $|f_3(\rho)|^2 \sim \rho^{|\nu|}|K_{-1}(\rho)|^2 = \rho^{|\nu|-2} \leq \rho^{0} < \rho^{-2}$, as $\rho\rightarrow 0$. Thus, $f_3(\rho)$ is square integrable.
	
	\item $|f_4(\rho)|^2 \sim \rho^{-|\nu|}|K_0(\rho)|^2 = \rho^{-|\nu|}(\log \rho)^{2} \geq \rho^{-2}(\log \rho)^{2} > \rho^{-2}$, as $\rho\rightarrow 0$. Thus, $f_4(\rho)$ is NOT square integrable.
\end{itemize}

When $J=J'\neq \frac{pn+1}{2}$ and $J=J'\neq \frac{pn-1}{2}$, we find that
\begin{itemize}
	\item $|f_1(\rho)|^2\sim \rho^{-|\nu|-|2l-(pn+1)|}$: To make $|f_1(\rho)|^2<\rho^{-2}$ as $\rho\rightarrow 0$, we require
	\bea
	0 \leq |J-\frac{(pn+1)}{2}| < 
	\frac{-|\nu|}{2}+1\leq 0,
	\eea
	which is impossible.
	
	\item $|f_2(\rho)|^2 \sim \rho^{|\nu|-|2l-(pn-1)|}$: To make $|f_2(\rho)|^2<\rho^{-2}$ as $\rho\rightarrow 0$, we require
	\bea
	pn-l-\frac{3}{2} < J < l + \frac{1}{2}.
	\eea

	\item $|f_3(\rho)|^2 \sim \rho^{|\nu|-|2l-(pn+1)|}$: To make $|f_3(\rho)|^2<\rho^{-2}$ as $\rho\rightarrow 0$, we require
	\bea
	pn-l-\frac{1}{2} < J < l + \frac{3}{2}.
	\eea
	
	\item $|f_4(\rho)|^2\sim \rho^{-|\nu|-|2l-(pn-1)|}$: To make $|f_4(\rho)|^2<\rho^{-2}$ as $\rho\rightarrow 0$, we require
	\bea
	0 \leq |J-\frac{(pn-1)}{2}| < 
	\frac{-|\nu|}{2}+1\leq 0,
	\eea
	which is impossible.
\end{itemize}
Therefore, $c_{1}$ and $c_{4}$ have to be zero to make the zero mode wavefunction square integrable. In addition, we would require $ pn-l-\frac{1}{2} < J < l+\frac{1}{2}$ to make $f_1(\rho)$ and $f_4(\rho)$ normalizable. To summarize, the zero mode wavefunction of the $Z_n$ vortex takes the following form:
\begin{align}
\Psi_{J,p,l}(\rho,\theta)=\frac{\rho^{\frac{pn}{2}-l}}{\sqrt{{\cal N}}}\begin{pmatrix}
	0 \\
	e^{i(J-l+\frac{1}{2})\theta}K_{J-\frac{pn-1}{2}}(\rho) \\
	-ie^{i(J-pn+l-\frac{1}{2})\theta}K_{J-\frac{pn+1}{2}}(\rho)\\
	0
\end{pmatrix}
\end{align}
where ${\cal N}$ is the normalization factor. This solution is physical only when 
\bea
pn-l-\frac{1}{2} < J < l+\frac{1}{2},
\eea
provided that $\nu = pn-2 l <0$.

\section{Bosonization analysis of the 1D vortex-line modes}
\label{app:1dluttinger}
Here we show that a group of vortex-line modes can always be symmetrically gapped out if the net vorticity vanishes. Consider a generic 1D nonchiral theory with the considered relevant symmetries.
At the single-particle level, it always has $4\N$ 1d modes at $+$ valley and $4\N$ 1D modes at $-$ valley, where $\N$ is not the same as the number $N$ of 3D Dirac points at one valley since the number of 1d modes also relies on the vortex windings.
We can split the $8\N$ modes into $\N$ groups, with each group furnishing irreps of the relevant symmetries.
In the following, we focus on one generic group of $8$ modes.

Given any group of $8$ modes, we can label the corresponding fermionic fields as
\eq{
(\psi_{1,R},\psi_{2,R},\psi_{3,R},\psi_{4,R};\psi_{1,L},\psi_{2,L},\psi_{4,L},\psi_{4N,L})\ ;
}
or equivalently, we can define the corresponding boson field vector as
\bea
{\bf \Phi}=(\phi_{1,R},\phi_{2,R},\phi_{3,R},\phi_{4,R};\phi_{1,L},\phi_{2,L},\phi_{3,L},\phi_{4,L})^T.
\eea
The Abelian bosonization convention we are following is
\bea
\psi_{L} \sim e^{i\phi_{L}},\ \psi_{R} \sim e^{-i\phi_{R}}.
\eea
Here we have omitted the Klein factors for simplicity. $\phi_{L}$ and $\phi_{R}$ are the chiral boson fields. The corresponding K-matrix is 
\bea
{\cal K}=\begin{pmatrix}
	\mathbb{I}_{4} & 0 \\
	0 & -\mathbb{I}_{4} \\
\end{pmatrix}.
\eea
The symmetry constraints of ${\bf \Lambda}_i$ are listed as follows:

\begin{itemize}

	\item $U(1)_c$: The charge vector ${\bf t}_c$ is found to be 
	\bea
	{\bf t}_c=(1,1,1,1;1,1,1,1)^T
	\eea 
	
	\item $\P$: 
	\eq{\P \bsl{\Phi} \P^{-1} =  \sigma_x\otimes \sigma_0 \otimes \sigma_x \bsl{\Phi} }
	
	\item $T_z$: Without loss of generality, we assume that the Dirac fermions with the channel index $v$ lives at $k=v k_0$. Then, the lattice translation symmetry $T_z$ leads to the valley charge vector ${\bf t}_v$:
	\bea
	{\bf t}_v= (1,-1,1,-1;1,-1,1,-1)^T
	\eea 
	\item $C_n$: 
	In the Luttinger liquid language, the $C_n$ symmetry will transform the fermionic modes in the following way:
	\bea
	 \ && C_n \psi_{1,R} C_n^{\dagger} = e^{\ii\frac{2\pi}{n}j_1} \psi_{1,R} \nonumber \\
	&&C_n \psi_{1,L} C_n^{\dagger} = e^{\ii\frac{2\pi}{n}j_2} \psi_{1,L}; \nonumber \\
	 \ && C_n \psi_{2,R} C_n^{\dagger} = e^{\ii\frac{2\pi}{n}j_2} \psi_{2,R} \nonumber \\
	&&C_n \psi_{2,L} C_n^{\dagger} = e^{\ii\frac{2\pi}{n}j_1} \psi_{2,L}; \nonumber \\
	 \ && C_n \psi_{3,R} C_n^{\dagger} = e^{-\ii\frac{2\pi}{n}j_1} \psi_{3,R} \nonumber \\
	&&C_n \psi_{3,L} C_n^{\dagger} = e^{-\ii\frac{2\pi}{n}j_2} \psi_{3,L}; \nonumber \\
	 \ && C_n \psi_{4,R} C_n^{\dagger} = e^{-\ii\frac{2\pi}{n}j_2} \psi_{4,R} \nonumber \\
	&&C_n \psi_{4,L} C_n^{\dagger} = e^{-\ii\frac{2\pi}{n}j_1} \psi_{4,L}.
	\eea
	Then, in the boson basis,
	\bea
	C_n  \bsl{\Phi}  C_n^{\dagger} =  \bsl{\Phi}  - \frac{2\pi}{n}   {\bf t}_n, 
	\eea
	where we have defined
	\bea
	{\bf t}_n&= (j_1,j_2,-j_1,-j_2;j_2,j_1,-j_2,-j_1).
	\eea
	
	\item $\mathcal{PT}$: 
	\eq{
	\mathcal{PT} \bsl{\Phi}  \mathcal{PT}^{\dagger} = - \sigma_0\otimes \sigma_x \otimes \sigma_0 \bsl{\Phi} + \frac{\pi}{2} \bsl{t}_{\PT}
	}
	with $\bsl{t}_{\PT}=(1,1,-1,-1;1,1,-1,-1)^T$.
\end{itemize}
Given the above symmetry transformation of the boson field vector, we find that the gapping vector ${\bf \Lambda}$ (which is always real) and its corresponding ${\bf \Phi}_{\Lambda}= {\bf \Lambda}^T \bsl{\Phi} $ must satisfy the following symmetry constraints to preserve required symmetries:
\eqa{
& U(1)_c: {\bf \Lambda}^T {\bf t}_c = 0  \\
& T_z:{\bf \Lambda}^T {\bf t}_v = 0  \\
& C_n: {\bf \Lambda}^T {\bf t}_n = 0\ (\text{mod } 2n) \\
& \P: \cos {\bf \Phi}_{\Lambda} = \cos (\bsl{\Lambda}^T\sigma_x\otimes \sigma_0 \otimes \sigma_x {\bf \Phi} ) \\
& \PT: \cos {\bf \Phi}_{\Lambda} = \cos (\bsl{\Lambda}^T\sigma_0\otimes \sigma_x \otimes \sigma_0 {\bf \Phi}- \frac{\pi}{2} {\bf \Lambda}^T  \bsl{t}_{\PT}).
\label{Eq: Sym Constraint of Class B}
}
In addition, ${\Lambda}$ needs to be checked to avoid SSB and to satisfy the null vector condition.

Based on the above equation, there are 4 linearly independent $\Lambda$'s that preserve $U(1)_c$, $T_z$ and $C_n$ as
\bea
{\bf \Lambda}_1 &=& (1,1,0,0;-1,-1,0,0)^T, \nonumber \\
{\bf \Lambda}_2 &=& (0,0,1,1;0,0,-1,-1)^T, \nonumber \\
{\bf \Lambda}_3 &=& (1,-1,0,0;0,0,-1,1)^T, \nonumber \\
{\bf \Lambda}_4 &=& (0,0,1,-1;-1,1,0,0)^T.
\label{Gapping vectors for Class B}
\eea
Furthermore, $\PT$ requries 

\bea
\langle {\bf \Phi}_{\Lambda_1} \rangle = \langle {\bf \Phi}_{\Lambda_2} \rangle,\ \ 
\langle {\bf \Phi}_{\Lambda_3} \rangle = \langle {\bf \Phi}_{\Lambda_4} \rangle\ ,
\label{Eq: Symmetry req. Z_2 Class B}
\eea
where as the inversion symmetry does not give any extra constraints.
%
Therefore, we have shown the existence of 4 linearly indepedent gapping vectors for each group of $8$ 1d modes.

Next, we show those gapping vectors do not lead to SSB. If the SSB of $C_n$, $\P$ or $\PT$ symmetry happens, there necessaily exists a {\bf non-zero} $C_n$/$\P$/$\PT$-breaking order parameter $\Delta$. Generally, $\Delta$ is defined as the expectation value of some vortex operator:
\bea
\Delta=\langle \Omega | e^{i{\bf L}_{\Delta}\cdot {\bf \Phi}} |\Omega \rangle\ ,
\eea
where $|\Omega \rangle$ is any ground state.
In fact, $\Delta\neq 0$ is possible if and only if ${\bf L}_{\Delta}$ can be linearly expanded in terms of the gapping vectors ${\bf \Lambda}_i$s:
\bea
{\bf L}_{\Delta}=\sum_{i=1}^{4} c_i{\bf \Lambda}_i, 
\label{Eq: Class B Z2 non-zero condition}
\eea
where $c_i$ are some coeffients.
Otherwise, $e^{i{\bf L}_{\Delta}\cdot {\bf \Phi}}$ will fluctuate and forces $\Delta = 0$. However, ${\bf \Lambda}_i^T{\bf t}_n=0$ implies
\bea
{\bf L}_{\Delta}^T {\bf t}_n =0,
\eea
which is independent of our choice of $j_1$ and $j_2$ for ${\bf t}_n$. Thus, any non-zero $\Delta$ constructed from ${\bf \Lambda}_i$s must respect $C_n$ symmetry. In other words, all $C_n$-breaking orders are vanishing in the presence of ${\bf \Lambda}_i$s.

Similarly, the possibility of a $\P$/$\PT$-breaking order parameter can be ruled out. By applying the symmetry requirement from \eqnref{Eq: Symmetry req. Z_2 Class B}, it is easy to prove 

\eqa{
& \bra{\PT \Omega } e^{i{\bf L}_{\Delta}\cdot {\bf \Phi}} \ket{\PT \Omega } = \Delta
 \\
& \bra{\P \Omega } e^{i{\bf L}_{\Delta}\cdot {\bf \Phi}} \ket{\P \Omega } = \Delta,
}
when the non-zero condition in \eqnref{Eq: Class B Z2 non-zero condition} is satisfied.

Additionally, the Mermin-Wager theorem ensures that there is no SSB of $U(1)_c$. Since the lattice transition along $z$, $T_z$, is equivalent to the valley $U(1)$ at the low-energy, the Mermin-Wager theorem ensures that there is no SSB of $T_z$. Therefore, we can conclude that ${\Lambda}_i$s do not break any of the relevant symmetries.
Thus, each group of $8$ modes can be gapped out in the a symmetry-preserving way, so is the whole system.



\section{Review of the crystalline gauge fields}
\label{app:cgaugefield}
Here we review the definition of crystalline gauge fields, which is introduced by Thorngren and Else~\cite{Else2018} (see also Ref.~\onlinecite{Else2018, Nissinen2018tetrads, nissinen2020field, Song2021polarization, Gioia2021, Naren2021,Julian2021,Huang2022,Julian2022}). Recall that a usual gauge field for a internal symmetry is defined as a map $A:M \rightarrow BG$, where $BG$ is the classifying space for a internal symmetry group $G$. For crystalline symmetry, this definition needs to be generalized since it's usually believe that a crystalline symmetry acts as a combination of internal symmetry and a (subgroup of the) isometry group action on the underlying manifold, and the map $A$ only captures the first part. Thorngren and Else proposed that, when $G$ contains a crystalline symmetry, we should replace the classifying space by a space called the homotopy quotient $X//G$. The space $X//G$ has a concrete construction called a Borel construction (or Borel space), which is defined as $(X \times EG)/G$, where G acts diagonally on the product space $X \times EG$. In practice, we can just define $X//G := (X \times EG)/G$~\footnote{It's called a homotopy quotient because it behaves nicely under homotopy. Given two G-spaces $X$, $Y$ and a homotopy equivalence $f:X \rightarrow Y$ with $f$ being equivariant, the induced map $X//G \rightarrow Y//G$ will be a homotopy equivalence.}.

Here are some known facts about $X//G$.
\begin{itemize}
    \item 
          When $G=1$, $X//G = X$.
    \item 
          When $X=pt$, $X//G \cong BG$.
    \item
          When $X=\mathbb{R}^{d}$ (or any contractible space), $X//G \cong BG$
    \item
         When $G$ actoin is free, $X//G \cong X/G$.
\end{itemize}

Now we review a simplicial complex construction of $X//G$. The $0$-simplices of $X//G$ are given by
the elements $x \in X$. The $1$-simplices $x \rightarrow x'$ are given by the elements $g \in G$ for which $g \cdot x = x'$. A 2-simplex is added for every triple $g_{1}$, $g_{2}$, $g_{3}$ with $g_{1}g_{2} = g_{3}$. Higher simplices are also added for all higher relations in the group. A crystalline gauge field is defined as a map $\alpha: M \rightarrow X//G$, where, most of the time, we choose $X = \mathbb{R}^{d}$. In this case, there is actually a homotopy equivalence: $\mathbb{R}^{d}//G \cong BG$.

A simple example is given by the discrete translation symmetry $\Gamma \cong \mathbb{Z}$ in 1d. We choose $X = \mathbb{R}$. Since translation $\Gamma$ acts freely on $\mathbb{R}$, the homotopy quotient $X//\Gamma \cong X/\Gamma \cong S^{1}$. 

Let $M=\mathbb{R}$, there is a G-CW complex decomposition where lattice sites are 0-cells and unit cells are 1-cells. Now consider a dual cell decomposition where the dual 0-cells are located at the center of the unit cell. We can choose the map $t:M \rightarrow \mathbb{R}//\mathbb{Z} \cong S^{1}$ such that $t(m_{n}) = x_{0}$, where $m_{n} \in M$ is a vertex in the dual cell complex (center of the unit cell) and $x_{0} \in \mathbb{R}//\mathbb{Z}^{T} \cong S^{1}$ is the based point in $S^{1}$. A dual 1-cell automatically carries a $g$ label and in this case it's labeled by $1 \in \mathbb{Z}$ since a dual 1-cell is mapped to a 1-simplex $x \rightarrow x'$. We see that the translation gauge field essentially counts the number of sites. We note that the choice of the space $X//G$ is not unique as long as they are homotopy equivalent. In this example, there is a physical choice of the space $X//\Gamma \cong S^{1}$. We can interpret $X$ as the momentum space and the homotopy quotient $X//\Gamma \cong S^{1}$ is precisely the brillouin zone. The above map $t$ assign a momenta $2\pi n$ to each dual 0-cell labeled by $n$, where $n$ is an integer, and assign the momentum difference (which must be integer multiple of the reciprocal lattice vector) between two neighboring unit cell to each dual 1-cell $(n,n+1)$. The translation gauge field $t$ we choose then satisfies the following property
\begin{equation}
    \int_{P_{x}} t = 2\pi L,
\end{equation}
where $P_{x}$ is a 1-cycle across the whole system and $L$ is the number of lattice point. The translation gauge field defined in this way is essentially the same as the elasticity tetrad or vielbein \cite{Nissinen2018tetrads, nissinen2020field, Else2021qc, Huang2022}.

One way to couple a low-energy Dirac theory to a crystalline gauge field is introducing the spatially dependent mass terms introduced in Ref.~\cite{Huang2022}. Let $\mathcal{P}$ be the space of parameter of the mass terms, the crystalline symmetry has a non-trivial actions on $P$ and generally it's possible to focus on a subspace of $P$ such that the crystalline symmetry acts in the same way as $\mathbb{R}^{d}$. We can then formally identify $P \cong X$ and construct the homotopy quotient $P//G$. Therefore, we have a realization of the crystalline gauge field as the map $\alpha: M \rightarrow P//G$ and, specifically, a crystalline gauge field in this representation is given by the pullback of a differential 1-form in the parameter space $P$.


\section{Quantization conditions of crystalline gauge fields}
Here we give more details about the quantization conditions and deformation classes of the translation and rotational gauge fields. 

The phase of the mass term $\phi(z)$ lives in the parameter space $P$ with a non-trivial translation symmetry action. As a result, $\phi$ actually lives in $P//\Gamma_{z} \cong S^{1}$, where $\Gamma_{z}$ denotes the group of $z$-translation. Therefore, the spatially dependent phase of the mass term in \eqnref{eq:zmass} is a map: $M \rightarrow P//\Gamma_{z} \cong B\Gamma_{z} \cong S^{1}$. The translation gauge field can then be defined explicitly as a differential 1-form $E_{z} = d\phi/2\pi$, where $\phi$ is the spatially dependent phase in the mass term \eqnref{eq:zmass}. We consider the configuration of $\phi$ such that the translation gauge field satisfies:
\begin{equation}
    \int_{P_z} E_{z} = 1,
\label{eq:Econd}
\end{equation}
where $P_{z}$ is a 1-cycle across one unit cell along the $z$-direction. The translation gauge field, satisfying \eqnref{eq:Econd} corresponds to the generators in the cohomology group $H^{1}(B \Gamma,\mathbb{Z}) = \mathbb{Z}$, which counts the number of lattice sites in the $z$-direction\cite{Nissinen2018tetrads, nissinen2020field, Else2021qc, Huang2022}. 

Similarly, the phase of the vortex mass term \eqnref{eq:vmass} gives a map: $M \rightarrow X//C_{n} \cong BC_{n}$. We then define the $C_{n}$ rotational gauge field as a differential 1-form $\omega^{(n)} = d n_\theta/2\pi$, where $n_{\theta} = \nu \theta$. We consider the configuration of $\theta$ such that the rotational gauge field satisfies: 
\begin{equation}
    \int_{P_n} \omega^{(n)} = \frac{\nu}{n} \ \text{mod} \ 1,
\label{eq:rot_cond}
\end{equation}
where $P_{n}$ is a 1-cycle with the starting and the points related by the $C_{n}$ rotation. The rotational gauge field is classified by the cohomology group $H^{1}( BC_{n},\mathbb{Z}) = \mathbb{Z}_{n}$. The 2-form $d\omega^{(n)}$ constructed by the rotational gauge field in the cohomology class in $H^{1}( BC_{n},\mathbb{Z})$ specified by \eqnref{eq:rot_cond} has to satisfy 
\begin{equation}
    \int_{D_{xy}} d\omega^{(n)} = \nu \ \text{mod} \ n,
\end{equation}
where $D_{xy}$ is any 2D disk containing the rotational axis along the $xy$-direction. This can be seen by the Stokes theorem:
\begin{align}
    \int_{D_{xy}} d\omega^{(n)}  &= \int_{\partial D_{xy}} \omega^{(n)} 
    \\
    &= n \int_{P_{n}} d\omega^{(n)} 
    \\
    &= \nu \ \text{mod} \ n,
\end{align}
where in the second equality we have used the fact that any 2D disk can be deformed into a disk respecting the $C_{n}$ rotation and the boundary of such disk is the union of the 1-cycle $P_{n}$.

\section{Classification details in the topological field theory}
\label{app:sol}
We now solve
\eq{
\label{eq:N_nu_n}
N \nu \mod n =0 \text{ and }N\in\dsZ
}
for $N$ under the assumption that
\eq{
\nu\in \dsZ-\{ 0 \},\ n\in\{1,2,3,...\}\ .
}
First, we note that \eqnref{eq:N_nu_n} is equivalent to
\eq{
N |\nu| \in n \dsZ \text{ and } N|\nu| \in |\nu| \dsZ\ ,
}
which is further equivalent to 
\eq{
N|\nu| \in |\nu| \dsZ \cap n \dsZ\ .
}
Therefore, the solution set to \eqnref{eq:N_nu_n} is 
\eq{
\frac{|\nu| \dsZ \cap n \dsZ}{|\nu|}\ .
}
Note that $|\nu| \dsZ \cap n \dsZ$ is the set of all common multiples of $|\nu|$ and $n$, which is further equal to the set of multiples of $\lcm(|\nu|,n)$, where $\lcm(|\nu|,n)$ is the least common multiple of $|\nu|$ and $n$~\cite{Garrett2007AA}.
It means that we have
\eq{
|\nu| \dsZ \cap n \dsZ = \lcm(|\nu|,n)\dsZ\ .
}
Owing to $\lcm(|\nu|,n) \gcd(|\nu|,n) = |\nu| n$~\cite{Garrett2007AA}, the solution set to \eqnref{eq:N_nu_n} eventually becomes 
\eq{
\frac{|\nu| \dsZ \cap n \dsZ}{|\nu|} = \frac{\lcm(|\nu|,n)\dsZ}{|\nu|} = \frac{|\nu| n}{\gcd(|\nu|,n) |\nu|}\dsZ = \frac{n}{\gcd(|\nu|,n) }\dsZ\ .
}

\end{document}